\documentclass[sn-nature]{sn-jnl}

\usepackage{graphicx}%
\usepackage{multirow}%
\usepackage{amsmath,amssymb,amsfonts}%
\usepackage{amsthm}%
\usepackage{mathrsfs}%
\usepackage[title]{appendix}%
\usepackage{xcolor}%
\usepackage{textcomp}%
\usepackage{manyfoot}%
\usepackage{booktabs}%
\usepackage{algorithm}%
\usepackage{algorithmicx}%
\usepackage{algpseudocode}%
\usepackage{listings}%
\usepackage{caption}
\usepackage{subcaption}

\theoremstyle{thmstyleone}%
\theoremstyle{thmstyletwo}%

\theoremstyle{thmstylethree}%

\begin{document}

\title[Article Title]{Echocardiogram Foundation Model - Application 1: Estimating Ejection Fraction}


\author*[1,2]{\fnm{Adil} \sur{Dahlan}}\email{adahlan@stanford.edu}
\equalcont{These authors contributed equally to this work.}

\author*[2]{\fnm{Cyril} \sur{Zakka}}\email{czakka@stanford.edu}
\equalcont{These authors contributed equally to this work.}

\author[3]{\fnm{Abhinav} \sur{Kumar}}

\author[4]{\fnm{Laura} \sur{Tang}}

\author[5]{\fnm{Rohan} \sur{Shad}}

\author[2]{\fnm{Robyn} \sur{Fong}}

\author*[2]{\fnm{William} \sur{Hiesinger}}\email{willhies@stanford.edu}

\affil*[1]{\orgdiv{School of Medicine}, \orgname{University College Dublin}}

\affil[2]{\orgdiv{Department of Cardiothoracic Surgery}, \orgname{Stanford Medicine}}

\affil[3]{\orgdiv{Department of Computer Science}, \orgname{Stanford University}}

\affil[4]{\orgdiv{Temerty School of Medicine}, \orgname{University of Toronto}}

\affil[5]{\orgdiv{Division of Cardiovascular Surgery}, \orgname{Penn Medicine}}


\abstract{
Cardiovascular diseases stand as the primary global cause of mortality. Among the various imaging techniques available for visualising the heart and evaluating its function, echocardiograms emerge as the preferred choice due to their safety and low cost. Quantifying cardiac function based on echocardiograms is very laborious, time-consuming and subject to high interoperator variability. In this work, we introduce EchoAI, an echocardiogram foundation model, that is trained using self-supervised learning (SSL) on 1.5 million echocardiograms. We evaluate our approach by fine-tuning EchoAI to estimate the ejection fraction achieving a mean absolute percentage error of 9.40\%. This level of accuracy aligns with the performance of expert sonographers.
}

\maketitle

\section{Introduction}\label{sec1}
The global burden of cardiovascular disease (CVD) has nearly doubled between 1990-2020 and is responsible for more than 20\% of all deaths in the United States \cite{CDCWONDER}. With a recent uptick in mortality rates in the last decade, current trends predict that nearly half of US adults are expected to have CVD by 2035\cite{dunbar_projected_2018}. Yet, despite the overall prevalence of heart disease, there exists well-established disparities in access to quality care and disease outcomes among the various genders, races, socioeconomic, and geographical groups affected. At the root of these inequalities, are unrecognized, and therefore untreated risk factors for heart disease, stemming from problems with provider bias, insurance coverage, access to care, and infrastructure \cite{havranek_social_2015}.

To alleviate some of the threats posed by these social determinants of health, there has been a rich line of works exploring the use of artificial intelligence (AI) in cardiovascular medicine \cite{ghorbani_deep_2020, ouyang_video-based_2020, zhou_artificial_2021}. At the forefront of these advances, applications of AI in echocardiography are uniquely-suited to narrow the gap in CV-related care, by enabling the safe and cost-effective assessment of cardiac anatomy and function in real-time, without some of the shortcomings traditionally associated with this imaging technique, such as inter-observer variability and artifacts \cite{feigenbaum_evolution_1996, cheitlin1997acc}. However, despite matching and sometimes even exceeding human experts in a variety of clinically-relevant tasks, these models have struggled to make headway in clinical adoption, due to key issues stemming from their lack of transparency and reliability \cite{shad_predicting_2021, rajpurkar_deep_2018, attia_artificial_2019, rajpurkar_appendixnet_2020}.

Most modern machine learning algorithms are characterized by the complex interplay of millions of artificial neurons, making it almost impossible to identify the underlying reasoning behind their predictions \cite{amann_explainability_2020}. By extension, it is not uncommon for these models to rely on medically irrelevant features and spurious correlations in their training data, leading to dangerously poor performances when deployed across hospitals, despite strong baselines in the lab \cite{cohen_problems_2021, damour_underspecification_2020, zhang_shifting_2022, chiang_youd_2021}. These models have also been found to be especially susceptible to mismatches between their training data and the data on which they are deployed, as a result of changes in patient demographics or data collection procedures \cite{eche_toward_2021, subbaswamy_development_2020, saria_tutorial_2019}. Failing to monitor and correct for these shifts can significantly degrade model performance and perpetuate their biases \cite{noseworthy_assessing_2020}. Lastly, these standard models are severely limited in terms of their downstream applicability, and only learn a narrow subset of visual concepts required to solve their specified task \cite{yu_external_2022}. In turn, this limits their clinical deployment, making it difficult to update or adapt them to new clinical diagnoses or other related tasks, without incurring significant cost and effort \cite{rajpurkar_ai_2022}.

As it stands, there exists a desperate need to overcome the shortcomings surrounding the transparency and reliability of current AI solutions. In a system that prides itself on evidence-based care, current approaches only serve to further erode the trust in automated medical decision-making processes and fail to fulfil their promise to narrow existing healthcare disparities. 

Video Vision Transformer (ViViT) models currently stand as the state-of-the-art for achieving optimal performance across a range of video-related tasks \cite{arnab_vivit_2021, dosovitskiy_image_2020}. However, these models exhibit a high demand for data and are prone to overfitting, even when trained on datasets exceeding one million inputs \cite{he_masked_2021}. This poses a significant challenge in the context of medical applications, where there is a shortage of labelled data. Adopting ViViT architectures in the medical domain becomes particularly challenging due to the limited availability of appropriately annotated datasets.

To mitigate the challenge of overfitting in medicine, various self-supervised learning (SSL) techniques have been developed \cite{spathis_breaking_2022, wang_review_2023, krishnan_self-supervised_2022, zhang_dive_2023, chowdhury_applying_2021, spathis_breaking_2022}. In 2022, Feichtenhofer et al. \cite{feichtenhofer_masked_2022} addressed this concern by developing the masked autoencoding (MAE) technique specifically for ViViT models. The MAE pretraining approach involves masking 90\% of the input video pixels and training the model to predict the original video. 

Echocardiography stands out as the preferred imaging technique for evaluating both the structure and function of the heart. Developing a foundational ViViT model specifically trained for echocardiograms holds the potential to not only boost overall model performance but also reduce the dependency on extensive labelled datasets for various downstream tasks. 

In this work, we introduce EchoAI, a foundation model for echocardiograms. We train EchoAI on a comprehensive dataset comprising 1.5 million echocardiograms. This training dataset encompasses a diverse range of adult and paediatric echocardiograms, including various types and views sourced from both public and Stanford-private datasets. We further fine-tune EchoAI on the EchoNet Dynamic dataset (\cite{ouyang_echonet-dynamic_nodate}) to estimate the ejection fraction (EF). Remarkably, Without any task-specific modifications, EchoAI demonstrates an impressive capability to estimate the ejection fraction, achieving a mean absolute error of 4.34 (9.40\%). This performance is noteworthy, especially considering the inherent challenges in ejection fraction estimation. Notably, the inter-user variability for this task is typically 13.5\% \cite{farsalinos_head--head_2015}. Furthermore, the process of calculating the ejection fraction involves the intricate steps of detecting individual heartbeats and manually segmenting the left ventricle at both end diastole and systole for each heartbeat, making it a highly time-consuming task.

\section{Previous Work}
Over the past decade, several AI models have been developed for cardiac imaging including classifying echocardiogram views \cite{madani_fast_2018}, diagnosing atrial fibrillation \cite{sanchez_de_la_nava_artificial_2021}, detecting valvular dysfunction \cite{vaid_multi-center_2023, nedadur_artificial_2022}, detecting cardiac amyloidosis \cite{goto_artificial_2021}, predicting post-operative right ventricular failure \cite{shad_predicting_2021}, screening for cardiac contractile dysfunction \cite{attia_screening_2019, yao_artificial_2021}, predicting and diagnosing heart failure \cite{miyashita_predicting_2023, yasmin_artificial_2021}, detecting atrial septal defect, dilated cardiomyopathy, hypertrophic cardiomyopathy, prior myocardial infarction \cite{liu_deep_2023}. However, these rely on supervised learning techniques that require huge amounts of labelled data. Weak supervised applications have also been developed, such as for classifying aortic valve malformations using cardiac MRI \cite{fries_weakly_2019}

Several self-supervised learning applications in medicine have emerged over the past few years, including training foundation models for ECG recordings interpretation \cite{lai_practical_2023}, medical image segmentation \cite{wang_annotation-efficient_2021, de_bruijne_hierarchical_2021, ouyang_self-supervised_2022}, medical image classification \cite{huang_self-supervised_2023, azizi_big_2021}, image quality enhancement and denoising \cite{eun_deep-learning-based_2020, eom_statistically_2023}

While previous attempts have been made to train echocardiogram foundation models \cite{christensen_multimodal_2023}, our contributions are the following:
\begin{itemize}
    \item EchoAI is the only echocardiogram model that incorporates a video-based encoder trained using self-supervised learning.
    \item EchoAI is trained on 1.5 million echocardiograms, the largest dataset ever used in training an echocardiogram model.
\end{itemize}

\section{Results}
\subsection{Echocardiogram Foundation Model}
\begin{figure}[H]
\begin{subfigure}{.5\linewidth}
\centering
\includegraphics[width=6cm]{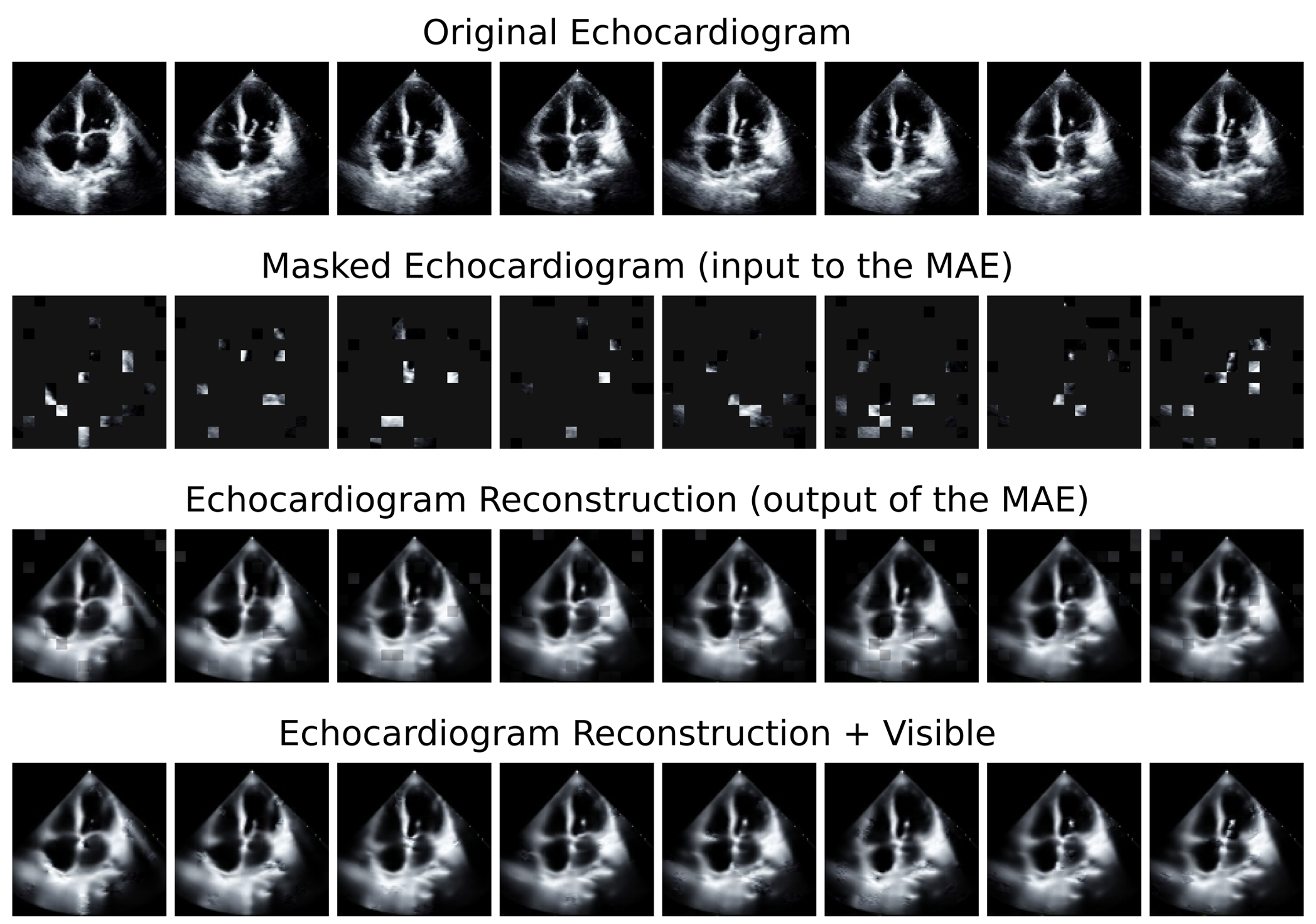}
\end{subfigure}%
\begin{subfigure}{.5\linewidth}
\centering
\includegraphics[width=6cm]{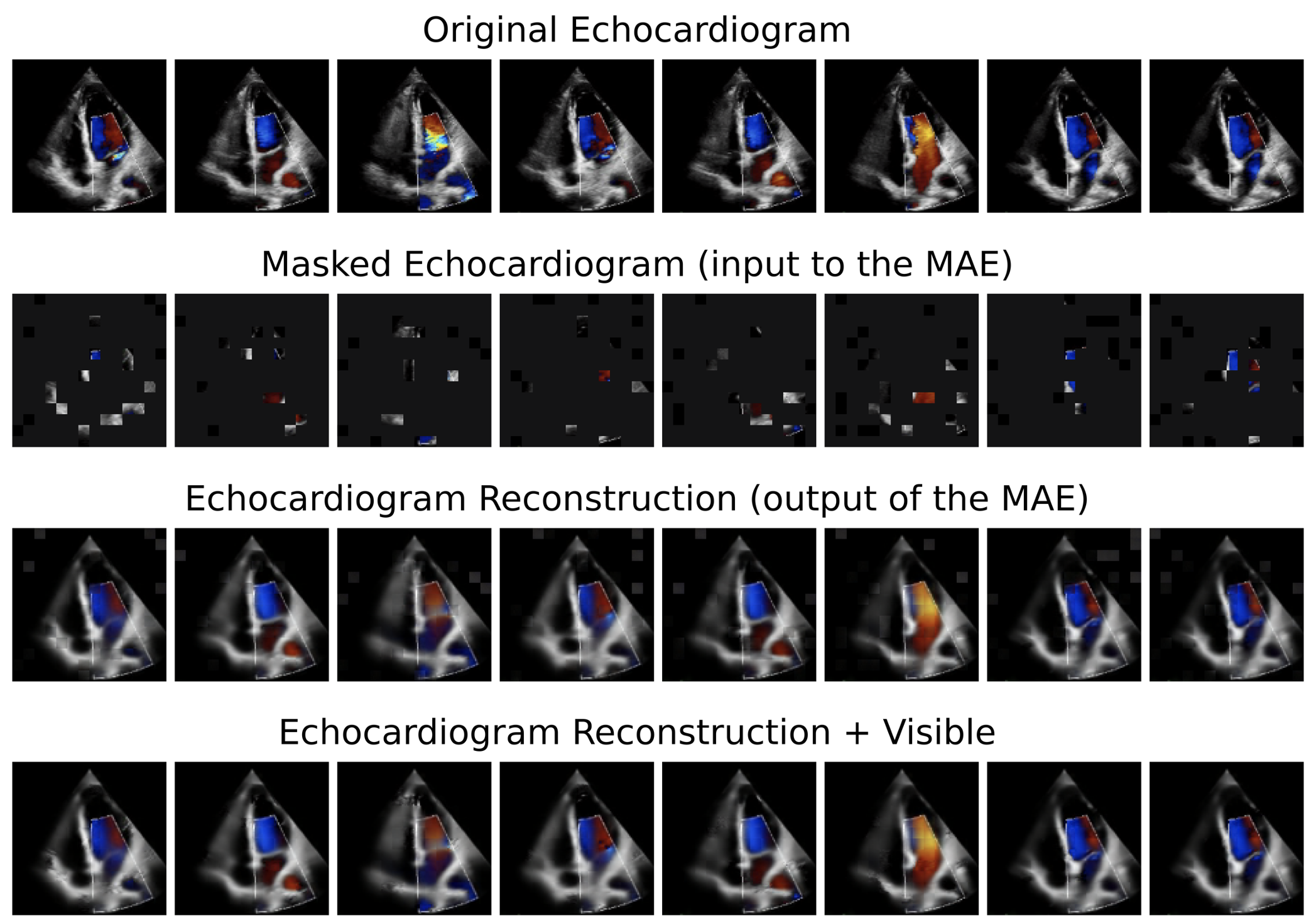}
\end{subfigure}
\caption{Two Original, Masked, Reconstructed and Reconstructed + Visible Echocardiograms output by EchoAI Masked Autoencoder}
\label{fig:MAE}
\end{figure}

We pretrain EchoAI using the masked autoencoding (MAE) self-supervised learning technique. In this approach, we train the model to reconstruct the original echocardiogram recording while exposing it to only 10\% of the video pixels. Following MAE pretraining, EchoAI demonstrates the capability to reconstruct various types and views of both adult and paediatric echocardiograms. We illustrate this process in Figure \ref{fig:MAE}.

\subsection{Estimating Ejection Fraction}
We finetune EchoAI using the EchoNet Dynamic datasets (\cite{ouyang_echonet-dynamic_nodate}) to estimate the ejection fraction. We evaluate the model's performance on the test datasplit comprising cases that were not part of the initial training dataset. The results showed a mean absolute error of 4.34 (9.40\%), a root mean squared error of 5.76 (15.02\%), and an R2 value of 0.78 when compared to the ground truth EF values measured by human experts. For a detailed analysis of the EF predictions, we provide the scatter plot in Figure \ref{Scatter_plot} and the corresponding performance metrics in Table \ref{Performance_metrics}.

\begin{minipage}{\textwidth}
\begin{minipage}[b]{0.49\textwidth}
\centering
\begin{tabular}{|l|l|}
\hline
Parameter                                                                            & Value   \\ \hline
\begin{tabular}[c]{@{}l@{}}Root Mean Squared\\ Error (RMSE)\end{tabular}             & 5.76    \\ \hline
\begin{tabular}[c]{@{}l@{}}Mean Absolute \\ Error (MAE)\end{tabular}                 & 4.34    \\ \hline
\begin{tabular}[c]{@{}l@{}}Root Mean Squared\\ Percentage Error (RMSPE)\end{tabular} & 15.02\% \\ \hline
\begin{tabular}[c]{@{}l@{}}Mean Absolute \\ Percentage Error (MAPE)\end{tabular}     & 9.40\%  \\ \hline
R2                                                                                   & 0.78    \\ \hline
\end{tabular}
\captionof{table}{Performance Metrics of EchoAI in Estimating Ejection Fraction on the EchoNet Dynamic Test Dataset}
\label{Performance_metrics}
\end{minipage}
\hfill
\begin{minipage}[b]{0.49\textwidth}
\centering
\includegraphics[width=6.4cm]{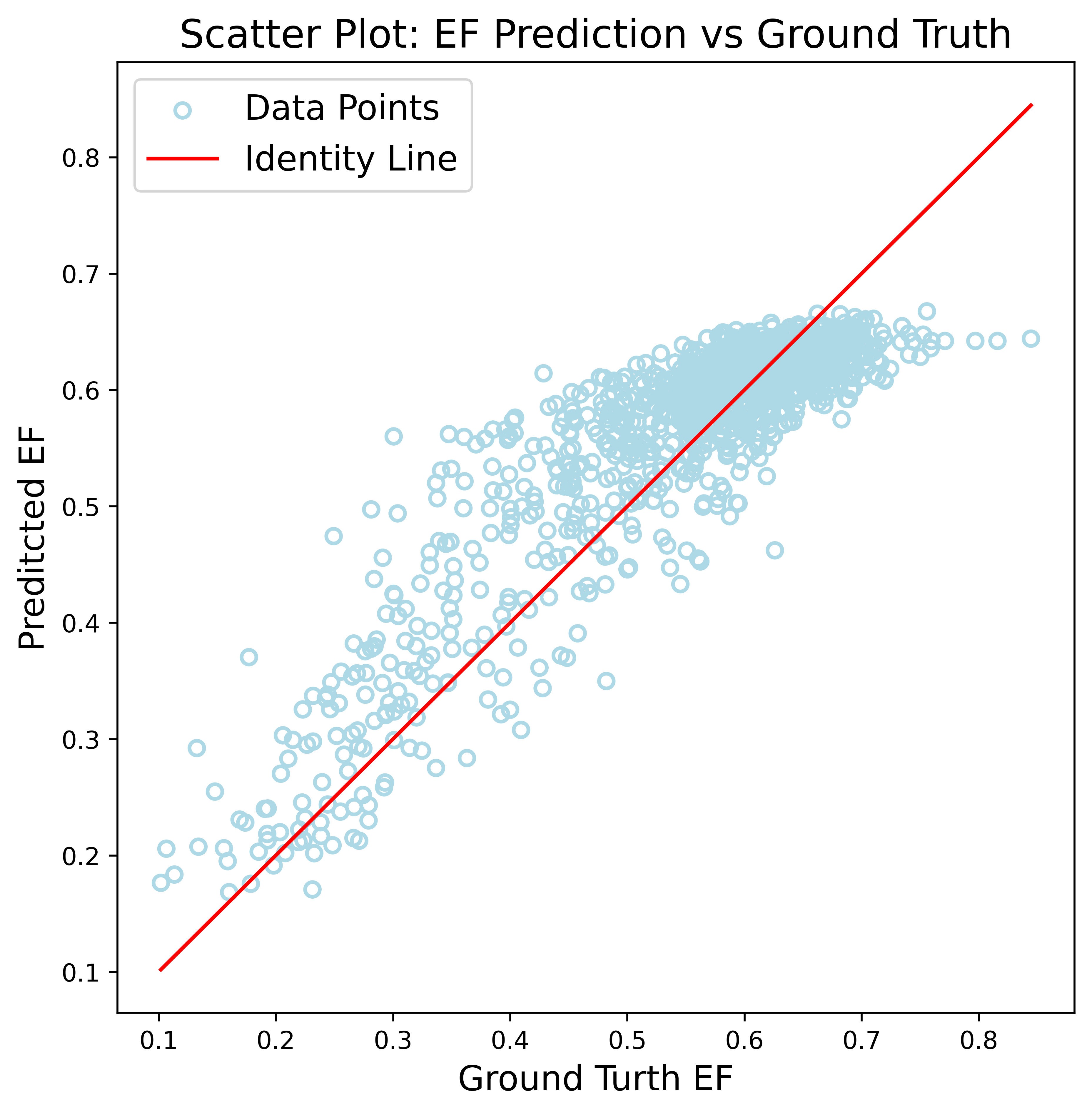}
\captionof{figure}{Scatter Plot of EF Estimation using EchoAI on the EchoNet Dynamic Test Dataset}
\label{Scatter_plot}
\end{minipage}
\end{minipage}

\section{Discussion}\label{sec4}
\subsection{Echocardiogram Foundation Model}
EchoAI distinguishes itself as an exceptionally robust echocardiogram foundation model, thanks to its extensive training on a large, diverse echocardiogram dataset. This dataset encompasses 1.5 million echocardiogram recordings, showcasing a variety of echocardiogram views including Apical 4 Chamber, Parasternal Short Axis, and Parasternal Long Axis. Additionally, it incorporates different types of echocardiograms, including Doppler, 2D, and 3D, providing a comprehensive understanding. Spanning various patient age groups, from adults to paediatrics, this broad training further enhances EchoAI's versatility and reliability as a robust foundation model for echocardiography.

\subsection{Estimating Ejection Fraction}
EchoAI demonstrates a mean absolute error of 4.34 (9.40\%) in estimating the ejection fraction, showcasing a level of performance comparable to human accuracy. It is worth noting that the inter-user variability in ejection fraction estimation stands at 13.5\% according to Farsalinos et al. (2015) \cite{farsalinos_head--head_2015}.

While EchoAI performance falls below the results reported in \cite{ouyang_echonet-dynamic_nodate}, where the mean absolute percentage error was 4.1\% and the root mean squared percentage error was 5.3\%, it is important to note that the approach used in \cite{ouyang_echonet-dynamic_nodate} involved the detection of 5 different heartbeats and the sampling of 32 frames from each heartbeat. This approach is computationally expensive and requires a significantly more complex task-specific implementation, and also requires labelling the end of systole and diastole by a medical professional.

Future work would involve training a multimodal echocardiogram foundation model as detailed in \cite{geng_multimodal_2022}. By means of incorporating meaningful information from both image and language, a multimodal echocardiogram foundation model trained on the echocardiogram recordings and the associated reports would be expected to have an enhanced performance.

\section{Conclusion}\label{sec5}
In this study, we introduce EchoAI, the pioneering foundation model for video-based echocardiograms. We train  EchoAI on a diverse dataset comprising 1.5 million echocardiogram videos, encompassing both adult and pediatric patients. Demonstrating its versatility, EchoAI serves as a potent foundation model for various downstream tasks related to echocardiograms. Notably, it exhibits superior performance, requiring minimal labelled data for fine-tuning.

Upon fine-tuning on the EchoNet Dynamic dataset, EchoAI excels in estimating the cardiac ejection fraction, achieving expert-level performance. The model attains a mean absolute error of 4.34 (9.40\%), showcasing its robust capability in enhancing diagnostic precision for echocardiogram-related analyses.

\textbf{Acknowledgments.} The computing for this project was performed on the Sherlock cluster. We would like to thank Stanford University and the Stanford Research Computing Center for providing computational resources and support that contributed to these research results.


\section{Methods}\label{sec2}
\subsection{Dataset}
We compile a total of 1,495,588 echocardiogram videos for training EchoAI, the echocardiogram foundation model. We source these echocardiograms from publicly available datasets and also from Stanford University Hospital, as detailed below:
\begin{enumerate}
    \item \textbf{EchoNet Dynamic}: comprising 10,030 Apical 4 Chamber echocardiograms of adult patients. This dataset is publicly available at \url{https://echonet.github.io/dynamic/}
    \item \textbf{EchoNet Paediatric}: comprising 7,810 echocardiograms of paediatric patients. These echocardiograms are split as follows: 3,285 are of the Apical 4 Chamber view and the remaining 4,527 are of the Parasternal Long Axis view. This dataset is publicly available at \url{https://echonet.github.io/pediatric/}
    \item \textbf{EchoNet LVH (Left Ventricular Hypertrophy)}: comprising 12,000 echocardiograms of adult patients. These echocardiograms are of the Parasternal Long Axis view. This dataset is publicly available at \url{https://echonet.github.io/lvh/}
    \item \textbf{Stanford Hospital Internal Dataset}: comprising 546,887 and 918,861 echocardiograms of adult and paediatric patients, respectively. The echocardiography DICOM files are sourced from the Department of Cardiothoracic Surgery at Stanford University (CA); (IRB 52440) with a waiver of consent owing to the retrospective nature of the research.
    
\end{enumerate}

A summary of the compiled datasets is in Table \ref{echo_datasets}. 

\begin{table}[h]
\caption{Echocardiogram datasets used for training EchoAI}\label{echo_datasets}%
\begin{tabular}{c cc c}
\hline
\multirow{2}{*}{Dataset} & \multicolumn{2}{c}{Number of Videos} & \multirow{2}{*}{Views} \\ \cmidrule{2-3}
& Adults & Paediatric & \\ \hline
EchoNet Dynamic \cite{ouyang_echonet-dynamic_nodate} & 10,030 & 0 & A4C \\ \hline
\multirow{2}{*}{EchoNet Paediatrics \cite{reddy_video-based_2023}} & \multirow{2}{*}{0} & \multirow{2}{*}{7,810} & A4C: 3,285 \\
& & & PSAX: 4,527 \\ \hline
EchoNet LVH \cite{duffy_high-throughput_2022} & 12,000 & 0 & PLAX\\ \hline
Stanford-Private & 546,887 & 918,861 & Miscellaneous \\ \hline
\end{tabular}
\end{table}

We adhere to a standardized data-preprocessing protocol for our videos. Initially, we resize the videos to align with the model's input dimensions. Additionally, we apply augmentations following the protocol outlined by Cubuk et al. (\cite{cubuk_randaugment_2019}). To ensure uniformity, we standardize the frame rates. Finally, we sample a specific number of frames at a designated rate or distribute them evenly throughout the entire video.

\subsection{EchoAI Training}
\begin{figure}[H]
    \centering 
    \includegraphics[width=12cm]{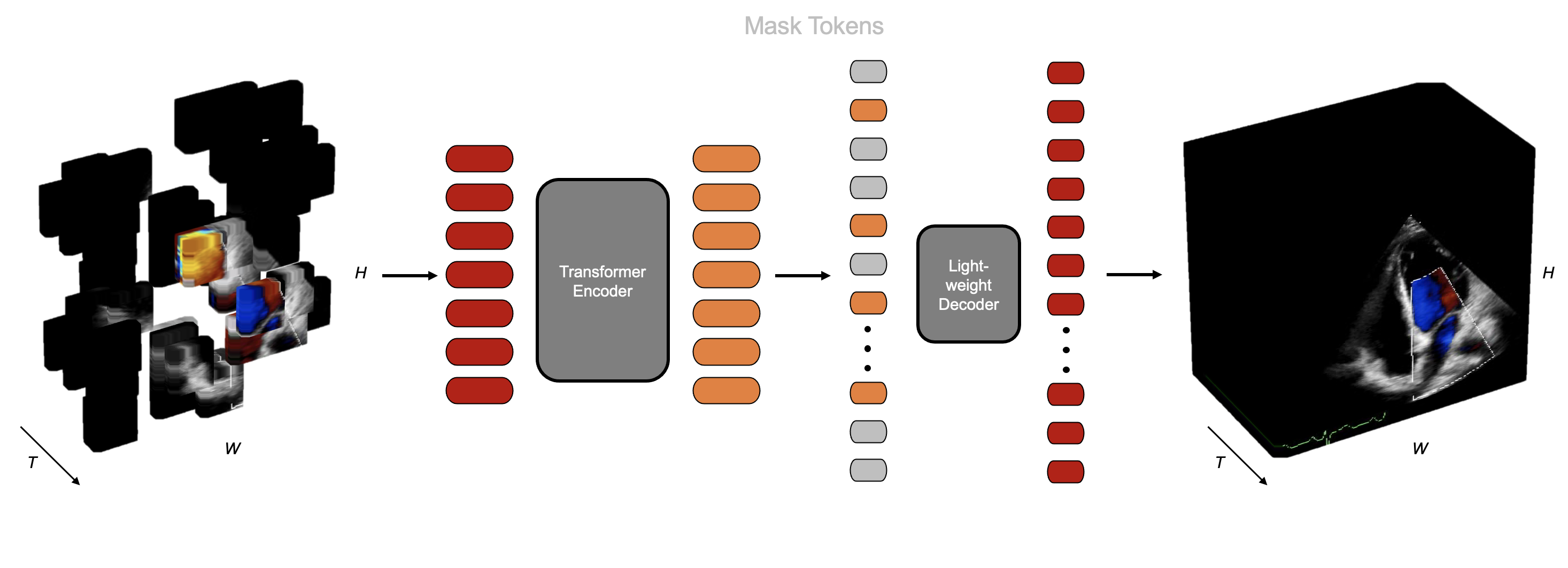} 
    \caption{EchoAI training pipeline - 90\% of patches are masked and the model is trained to reconstruct the original echocardiogram recording}
    \label{fig:MAE_ViViT_training_pipeline}
\end{figure}

EchoAI is a Video Vision Transformer model. We train EchoAI utilizing the Masked Autoencoding (MAE) technique, as elucidated in the study by Feichtenhofer et al. (2022) \cite{feichtenhofer_masked_2022}. The MAE training protocol entails the incorporation of a lightweight decoder into the video vision transformer model's encoder. The encoder is responsible for encoding the input into a single 2D matrix with specific dimensions, constituting the latent space representation of the input. Conversely, the decoder reconstructs the input image by utilizing the latent space representation, comprised of both visible tokens and mask tokens (refer to Figure \ref{fig:MAE_ViViT_training_pipeline}).

\subsection{Estimating the Ejection Fraction}
\begin{figure}[H]
    \centering 
    \includegraphics[width=12cm]{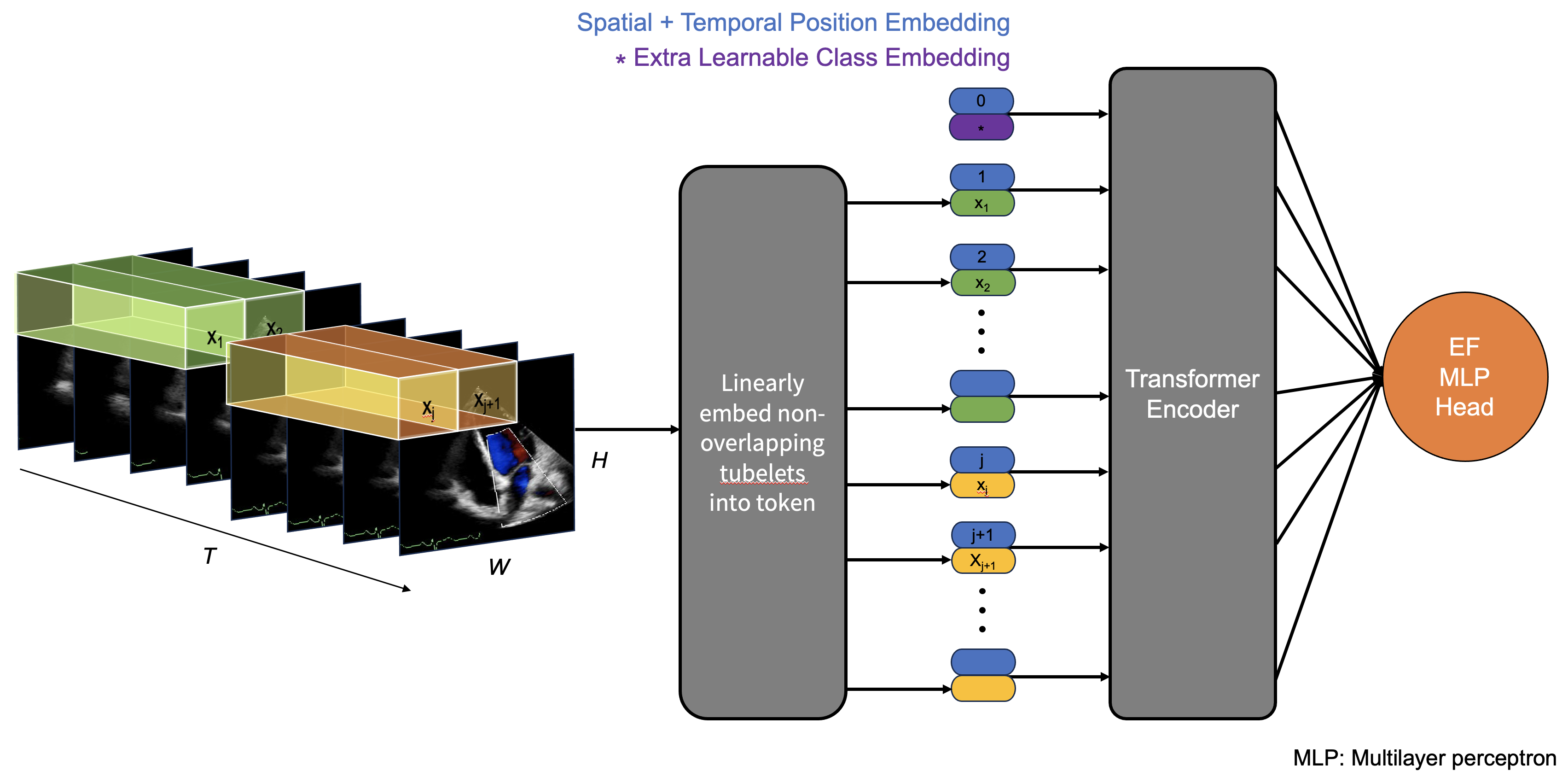} 
    \caption{EchoAI fine-tuning pipeline for estimating the ejection fraction on the EchoNet Dynamic test dataset}
    \label{fig:ViViT_Finetuning_pipeline_for_Estimating_the_EF}
\end{figure}

We fine-tune EchoAI to estimate the ejection fraction, as illustrated in Figure \ref{fig:ViViT_Finetuning_pipeline_for_Estimating_the_EF}. The fine-tuning protocol comprises several essential steps. Firstly, we preprocess the data and segment the input video into patches of specific dimensions. We then embed these input patches, transforming the 3D pixel space into a linear vector with learned values. To retain positional and temporal information, we add learnable positional and temporal embeddings to the embedded patches. Optionally, we add a  class token at the beginning of the sequence of embedded patches. Subsequently, we feed these embeddings through the transformer encoder, resulting in a latent representation of the initial image. Finally, we input the latent representation of the initial echocardiogram video through a dense layer which outputs the estimation of the ejection fraction.  Essentially, this constitutes a standard training pipeline for a video vision transformer designed for a regression task.

\subsection{EchoAI Training and Finetuning Pipeline}
We conduct training and fine-tuning of EchoAI over 50 epochs, distributing our code across four 40GB GPUs for parallel processing. Throughout this process, we employ the Mean Square Error (MSE) Loss, denoted as $MSE=\frac{1}{n}\sum_{i=1}^N(Y_i-\hat{Y}_i)^2$. This loss function serves to quantify the mean disparity between actual and predicted values, addressing both pixel values and ejection fraction estimation.

To optimize the learning process, we implement a standard cosine annealing scheduler, dynamically adjusting the learning rate per data iteration. The optimization is carried out using the AdamW optimizer, with base learning rates set at 0.0016 and 0.0024 for training and fine-tuning of EchoAI, respectively. 

To expedite training, we implement automatic mixed precision and accumulated gradients across 2 data iterations. For EchoAI pertaining and fine-tuning, we explore seven different hyperparameter configurations. These configurations are detailed in Appendix 1. To assess the significance of MAE pretraining, we also fine-tune a  "vanilla" (not pre-trained) ViViT model to predict the ejection fraction.

\section*{Declarations}
\subsection{Funding}
This project was supported in part by a National Heart, Lung, and Blood Institute (NIH NHLBI) grant (1R01HL157235-01A1) (W.H.) and by the AI in Medicine Scholarship from University College Dublin.

\subsection{Code availability}
The code and model weights for the EchoAI model are made available to the public at \url{https://github.com/adildahlan/echo_ssl}.




\pagebreak 
\section{Appendix 1}\label{app1}
\begin{table}[ht]
\caption{Hyperparameter search experiments for EchoAI training and fine-tuning on the EchoNet Dynamic dataset for estimating the ejection fraction}\label{table:HPSearchExperiments}
\begin{tabular}{cccccccc}
\hline
Experiment & \begin{tabular}[c]{@{}c@{}}Architecture\\ Size\end{tabular} & \begin{tabular}[c]{@{}c@{}}Image\\ Size\end{tabular} & \begin{tabular}[c]{@{}c@{}}Number of\\ Sampled\\ Frames\end{tabular} & \begin{tabular}[c]{@{}c@{}}Frames\\ Sampling\\ Rate\end{tabular} & \begin{tabular}[c]{@{}c@{}}Video\\ Target\\ fps\end{tabular} & \begin{tabular}[c]{@{}c@{}}Patch\\ Size\end{tabular} & \begin{tabular}[c]{@{}c@{}}Number of\\ Reconstructed\\ Frames\end{tabular} \\ \hline
1 & Large & 224x224 & 16 & 4 & 30 & 16x16 & 8 \\ \hline
2 & Base  & 112x112 & 32 & 3 & 50 & 16x16 & 32 \\ \hline
3 & Large & 112x112 & 32 & 3 & 50 & 16x16 & 32 \\ \hline
4 & Large & 112x112 & 16 & 4 & 50 & 16x16 & 8 \\ \hline
5 & Large & 112x112 & 16 & 4 & 50 & 16x16 & 8 \\ \hline
6 & Large & 112x112 & 32 & 3 & 50 & 16x16 & 8 \\ \hline
7 & Large & 112x112 & 32 & 3 & 50 & 16x16 & 16 \\ \hline
8 & Large & 112x112 & 32 & \begin{tabular}[c]{@{}c@{}}NA - equally \\ spaced frames\\  throughout\\ the video.\end{tabular} & 50 & 16x16 & 8 \\ \hline
\end{tabular}
\end{table}

\end{document}